\documentclass[runningheads]{llncs}
\usepackage{graphicx}
\usepackage[perpage,stable]{footmisc}
\usepackage{tikz}
\usepackage{comment} 
\usepackage{amsmath,amssymb} 
\usepackage{color}

\begin{document}
\pagestyle{headings}
\mainmatter
\def\ECCVSubNumber{100}  

\title{FAN: Frequency Aggregation Network for Real Image Super-resolution} 

\titlerunning{FAN}
%
\author{Yingxue Pang\footnotemark[1] \and Xin Li\footnotemark[1] \and  Xin Jin  \and \\ Yaojun Wu\and Jianzhao Liu \and Sen Liu \and Zhibo Chen\footnotemark[2]}
\authorrunning{Y. Pang et al.}
%
\institute{CAS Key Laboratory of Technology in Geo-spatial Information Processing and Application System,\\
University of Science and Technology of China, Hefei 230027, China
\email{\{pangyx,lixin666,jinxustc,yaojunwu,jianzhao\}@mail.ustc.edu.cn}\\
\email{elsen@iat.ustc.edu.cn}  \email{chenzhibo@ustc.edu.cn}}
\maketitle
\renewcommand{\thefootnote}{\fnsymbol{footnote}} 
\footnotetext[1]{The first two authors contributed equally to this work.} 
\footnotetext[2]{Corresponding author.} 
\begin{abstract}
Single image super-resolution (SISR) aims to recover the high-resolution (HR) image from its low-resolution (LR) input image. With the development of deep learning, SISR has achieved great progress. However, It is still a challenge to restore the real-world LR image with complicated authentic degradations. Therefore, we propose FAN, a frequency aggregation network, to address the real-world image super-resolu-tion problem. Specifically, we extract different frequencies of the LR image and pass them to a channel attention-grouped residual dense network (CA-GRDB) individually to output corresponding feature maps. And then aggregating these residual dense feature maps adaptively to recover the HR image with enhanced details and textures. We conduct extensive experiments quantitatively and qualitatively to verify that our FAN performs well on the real image super-resolution task of AIM 2020 challenge. According to the released final results, our team SR-IM achieves the fourth place on the X4 track with PSNR of 31.1735 and SSIM of 0.8728.
\keywords{Frequency aggregation network (FAN), Real image super-resoltion (RealSR), AIM 2020 challenge.}
\end{abstract}

\section{Introduction}
\begin{figure}[htp]
	\centering
	\includegraphics[width=\linewidth]{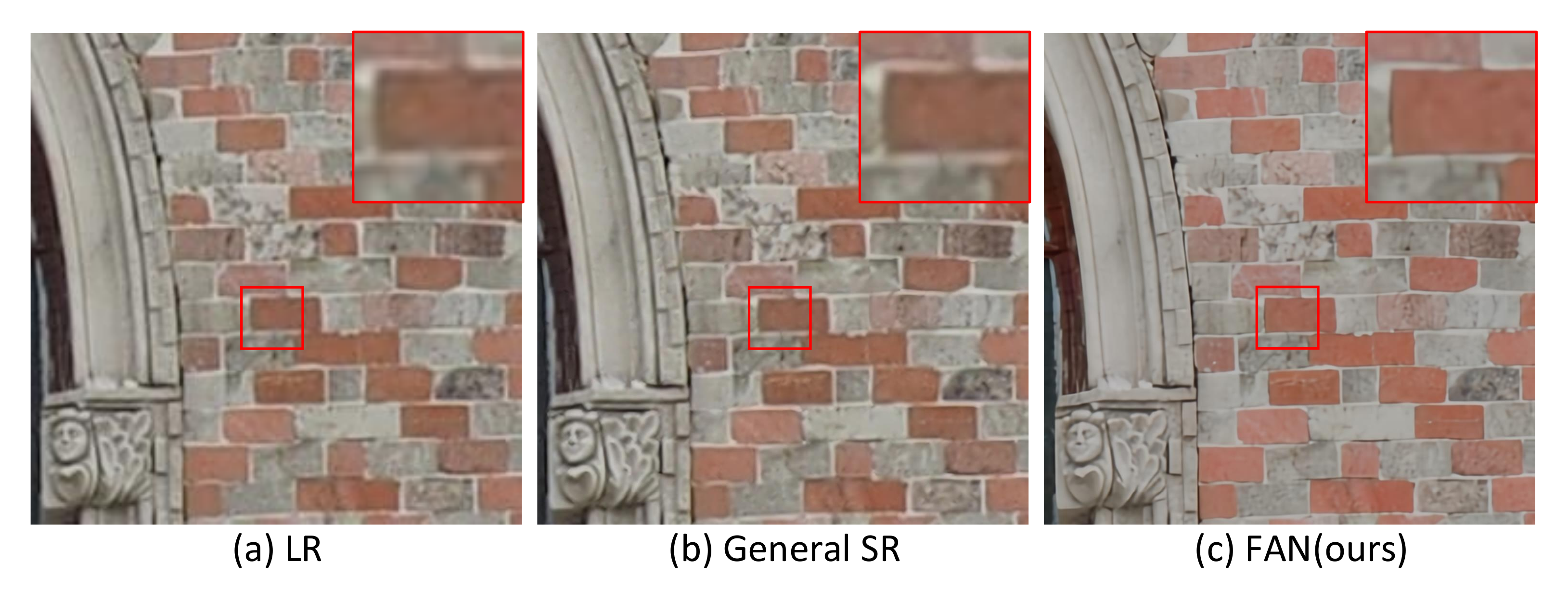}
	\caption{Examples of SISR cannot be generalized to RealSR. (a) Real low-resolution image, which is captured by low quality DSLR camera. (b) Processed by general SR model EDSR. (c) Processed by our FAN.}
	\label{fig:example}
\end{figure}

Single image super-resolution (SISR) task aims to recover the high-resolution (HR) image from its low-resolution (LR) input image, where the LR image is acquired by applying some downsamping settings to the HR image. 
Whether the early traditional methods like \cite{Freeman2002Example,chang2004super} or recent methods \cite{Dong2014Learning,Johnson2016Perceptual,Lai_2017_CVPR,Maeda_2020_CVPR}, SISR has drawn more and more attention because of its wide range of application, such as medical image\cite{greenspan2009super}, surveillance \cite{zhang2010super} and security \cite{gohshi2015real}.  
With the development of convectional neural networks (CNN) and several high quality super-resolution datasets, Recent models \cite{Dong2014Learning,Johnson2016Perceptual,Lai_2017_CVPR,Maeda_2020_CVPR,lan2020cascading,kalarot2020component} have achieved remarkable success in SISR task. A well-designed CNN can effectively capture the non-linear mapping and automatically learn the mapping function based on various high-quality super-resolution dataset. However, these models can perform well on the ``clean" standard benchmarks, and they often fail to be applied in real-world scenarios. They cannot handle the real-world image super-resolution problem with complicated degradation processes, as shown in Fig. \ref{fig:example}.

Therefore, the real-world image super-resolution (RealSR) has attracted more and more attention. Different from the general SISR, the purpose of RealSR is to learn a general model to restore LR image captured under practical scenarios to high-resolution. This model focuses on solving the degradation problem that is more complex than bicubic downsampling. The paired LR and HR images are captured by various DSLR cameras. The diverse degradation differences among devices can be expressed as Eq.\ref{degradation}
\begin{equation}
    \begin{split}
        & {{I}_{HR}}=(I*{{k}_{HR}})+{{n}_{HR}}, \\ 
 & {{I}_{LR}}=(I*{{k}_{LR}}){{\downarrow }_{\text{s}}}+{{n}_{LR}}, 
    \end{split}
\label{degradation}
\end{equation}
where ${k}_{LR}$ and ${k}_{HR}$ represent degradation kernel of high quality and low quality DSLR cameras respectively. ${\downarrow }_{\text{s}}$ denotes the multiple of downsampling. ${n}_{HR}$ and ${n}_{LR}$ are additive noise. $I$ represents the real world image. ${I}_{HR}$ and ${I}_{LR}$ are the image pairs collected with high-quality and low-quality DSLR cameras respectively. RealSR recovers the ${I}_{HR}$ from ${I}_{LR}$, which more satisfies the needs of industrial applications. According to Eq. \ref{degradation}, we can get the ${I}_{LR}$ from ${I}_{HR}$ as  Eq.\ref{LR_HR}.
\begin{equation}
        {{I}_{LR}}=({{I}_{HR}}*k_{HR}^{-1}*{{k}_{LR}}){{\downarrow }_{s}}+{{n}_{LR}}-({{n}_{HR}}*k_{HR}^{-1}*{{k}_{LR}}){{\downarrow }_{s}}.
\label{LR_HR}
\end{equation}
From the above equation, we can find that the RealSR is more complicated than general SISR. When the degradation of high-quality DSLR camera is small, the Eq. \ref{LR_HR} will be degraded to general SISR problem.

In this paper, we propose a frequency aggregation network (FAN) for complex real-world image super-resolution problem. Specifically, we extracted the low-frequency, middle-frequency and high-frequency of LR image with multi-scale representation. Then we utilize different branches to restore the corresponding frequency components of LR and aggregate them adaptively with channel attention mechanism to generate the HR image. From different frequency components, our framework can be more robust to solving different distortions. In order to solve the complex degradation kernels, we take RDB \cite{zhang2018residual} as our basic unit, and redesign GRDB \cite{kim2019grdn} into CA-GRDB by introducing a channel mechanism to adaptively aggregate the representations of different receptive fields, which can further improve the representation ability of the network. Extensive experiments have validated the effectiveness of our FAN for real image super-resolution problem. According to the released final results, our team SR-IM achieves the fourth place on the X4 track with PSNR of 31.1735 and SSIM of 0.8728 by applying our FAN on the real image super-resolution task of AIM 2020 challenge \cite{AIM2020_RSRchallenge,wei2020aim_realSR}.

Our contributions can be summarized as follows:
\begin{itemize}
\item We propose a frequency aggregation network (FAN) for real image-world super-resolution problem.
\item We redesign the novel CA-GRDB module by introducing a channel attention mechanism to further improve the representation ability of network.
\item According to the released final results, our team SR-IM achieves the forth place on the X4 track with PSNR of 31.1735 and SSIM of 0.8728 on the real-world image super-resolution task of AIM 2020 challenge \cite{AIM2020_RSRchallenge}.
\end{itemize}

\section{Related Works}
\subsection{Single Image Super-Resolution}
Dong et al. proposed the SRCNN \cite{Dong2014Learning,DongFSRCNN} which can be roughly seen as the first SISR work based on CNN. They adopted an end-to-end supervised learning model to restore the HR images with its corresponding bicubic downsamping LR images. Compared to traditional methods \cite{Freeman2002Example,chang2004super}, they reconstructed high-quality HR images with clearer details and higher metric scores. 
SRCNN proves the effectiveness of CNN in solving SISR problem and inspires plenty of works to be proposed to improve the qualitative and quantitative results. Kim et al. \cite{Kim_2016_CVPR} proposed DRCN in which they designed a deeper recursive layer with the receptive field of 41 to demonstrate their performance in common benchmarks. But methods extended by SRCNN \cite{Kim_2016_CVPR,tai2017image} usually utilize the pre-defined unsamping operator which increases unnecessary computational cost and lead to reconstruction artifacts in some cases. To tackle this problem, LapSRN \cite{Lai_2017_CVPR} constructed a set of cascaded sub-networks to progressively predicts the sub-band high frequency residuals. It replaced the pre-defined upsampling with the learned transposed convolutional layers to remove the undesired artifacts and reduce the computational cost.

Considered the methods mentioned above, we still have the difficulty to recover the HR image with more high-frequency details when using larger unsampling factors. Since those methods focus to optimize the pixel loss and ignore the image quality, the results often lack high-frequency details or fail to maintain the perceptual fidelity compared to its high-resolution ``ground truth". Ledig et al. \cite{ledig2017photo} proposed SRGAN, the first SISR work which utilized Generative Adversarial Network(GAN), to solve the problem. They combine a perceptual loss, an adversarial loss and a content loss to infer HR images for 4x upsampling factors and show hugely signiﬁcant gains in perceptual quality. Wang et al. \cite{wang2018esrgan} proposed ESRGAN which is an enhanced version by reconsidering the entire SRGAN and modifing it in network architecture, adversarial loss and perceptual loss. And ESRGAN brings the subjective
state-of-the-art algorithm to SR. Zhang et al. \cite{zhang2019ranksrgan} proposed RankSRGAN to introduce a ranker to learn the behavior of perceptual metrics and address the indifferentiable perceptual metrics problem.

Unlike previous feed-forward approaches, Haris et al. \cite{haris2018deep} introduced the back-projection into the reconstruction process of SISR. They showed that combining the up and down sampling, along with error feedbacks, encourages to get better results. Then, it seems to be a new trend to exploite the attention mechanism which can adaptively process visual information and focus on salient areas.  RCAN \cite{zhang2018image} was proposed to apply the channel attention to capture the dependencies among channels. RNAN \cite{zhang2019residual} utilized the local and non-local attention modules to get feature representation and dependencies.  Similarly, SAN \cite{dai2019second} also incorporated the non-local attention mechanism to capture long-range spatial contextual information.
In addition, multi-pass methods also prove its effectiveness in SISR in recent work \cite{dahl2017pixel,ren2017image,li2018multi,han2018image}. Within it, different paths may perform different operations to extract corresponding feature maps. By fusing these feature maps coming from each path, the whole network provides better modelling capabilities and generalization. We hence get inspired to construct our three-path aggregation network to capture different frequency infromation of LR images. And we demonstrate that our method performs well on the AIM 2020 challenge, came the fourth place on the X4 track with PSNR of 31.1735 and SSIM of 0.8728.

\subsection{Real-World Image Super-Resolution}
The SISR methods mentioned above usually utilize the clean and ideal datasets which mostly adopt simple and uniform degradation (e.g. bicubic downsampling) to construct LR images from HR images without any distracting artifacts (e.g. sensor noise, image compression, non-ideal PSF, etc). However, since the degradation in real world images is more complicated, SISR models inevitably fail in real-world image super-resolution with limited application and generalization.

To overcome the problem, several real-world super-resolution challenges \cite{lugmayr2019aim,lugmayr2020ntire} have been hold to attract more participants to come up their solutions. Cai et al. \cite{cai2019toward} build a real-world single super-resolution dataset where paired LR-HR images on the same scene are captured by adjusting the focal length of a digital camera.  Shocher et al. \cite{shocher2018zero} proposed ZSSR , the first unsupervised CNN-based SR method, to train a small image-specific CNN at test time by expoliting the internal recurrence information of the image. Fritsche et al. \cite{fritsche2019frequency} proposed DSGAN to generate LR-HR pairs with similar natural image characteristics.  Ji et al. \cite{ji2020real} proposed a new degradation framework with various blur kernels and noise injections to solve the realistic image super-resolution problem. However, in this work, we propose FAN to extract the different frequency components of the LR image and then aggregate them to recover the HR image with more high frequency details.

\section{Method}
In this section, we will illustrate the overall architecture of our FAN and explain each component in detail, including hierarchical feature extractor (HFE), the main body with three branches and fusion module.

\begin{figure}[htp]
	\centering
	\includegraphics[width=\linewidth]{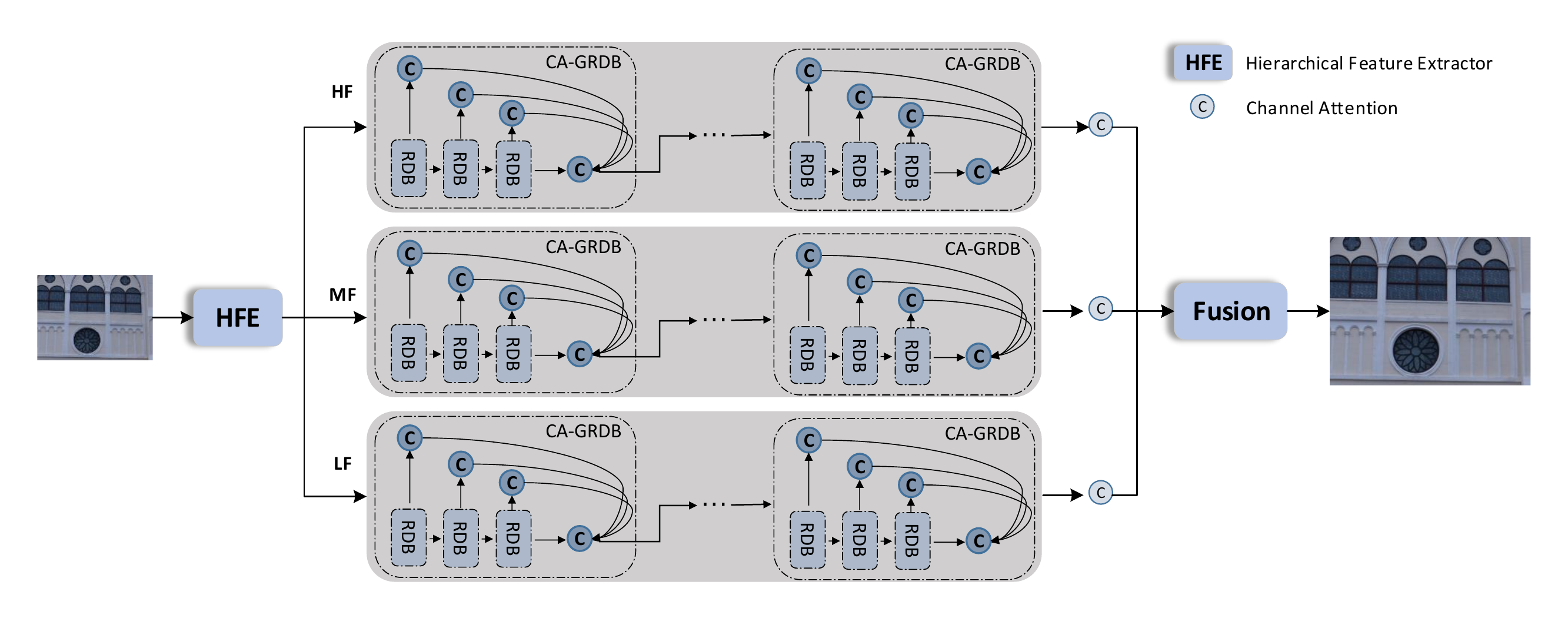}
	\caption{The proposed network architecture: FAN }
	\label{fig:framework}
\end{figure}

\subsection{Overall Structure of FAN}
Our end-to-end frequency aggregation network (FAN) is shown in Fig. \ref{fig:framework}. We first extract different frequency components of the LR image through the hierarchical feature extractor (HFE), and then pass them to each branch to obtain corresponding feature map. FAN contains three branches with same structure to process high-frequency component (HF), middle-frequency component (MF) and low-frequency component (LF) respectively. Followed three branches, FAN utilizes the fusion module to recover the HR image with enhanced details and textures from three residual dense feature maps.
\subsection{Hierarchical Feature Extractor}
\begin{figure}[htp]
	\centering
	\includegraphics[width=0.9\linewidth]{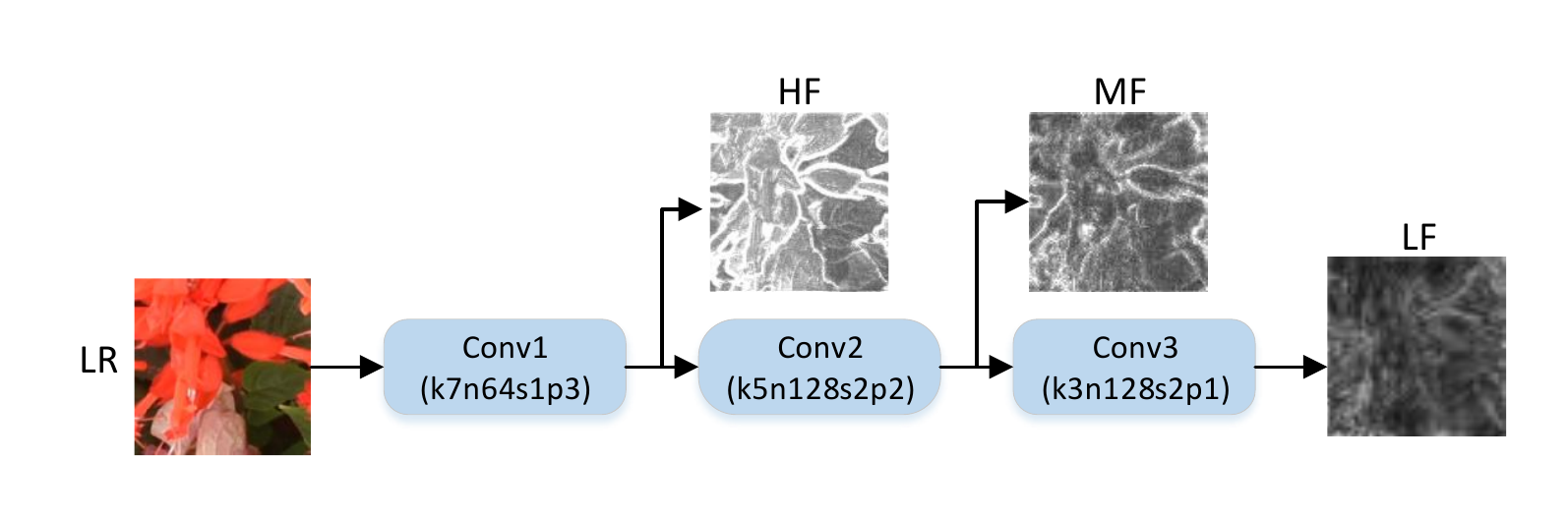}
	\caption{Hierarchical feature extractor (HFE) with corresponding kernel size (k), number of feature maps (n), stride (s) and padding size (p) for each convolutional layer.}
	\label{fig:FHE}
\end{figure}
Method \cite{chen2019drop} has proved that features extracted by different scale factors can capture different frequency information of the image. Inspired by it, we design the hierarchical feature extractor (FHE), which is shown in Fig. \ref{fig:FHE}, to progressively extract different frequency components.
Specifically, we gradually input the LR image into three convolutional layers and output the hierarchical feature maps in each layer. As \cite{chen2019drop} claimed, the larger feature map size is, the higher frequency component it contains. Therefore, we design the first convolutional layer (Conv1) with form of $k7n64s1p3$ to output the same size feature map (HF) as the LR image. The remaining two (Conv2 and Conv3) with forms of $k5n128s2p2$ and $k3n128s2p1$ produce the feature maps (MF and LF) which have half and quarter size of the input LR image. We indicate the convolutional layer by corresponding kernel size (k), number of feature maps (n), stride (s) and padding size (p).   
 Among three frequency components, LF contains the low-level frequency of LR image with roughly structure and HF represents the high-level frequency with fine-grained details such as edges and textures. MF indicates the intermediate value between them. The entire frequency component extraction process can also be defined as Eq. \ref{feature-define}.

\begin{equation}
    \begin{split}
         & HF=Conv1(I_{LR}), MF=Conv2(HF), LF=Conv3(MF)
    \end{split}
\label{feature-define}
\end{equation}

\begin{figure}[htp]
	\centering
	\includegraphics[width=0.9\linewidth]{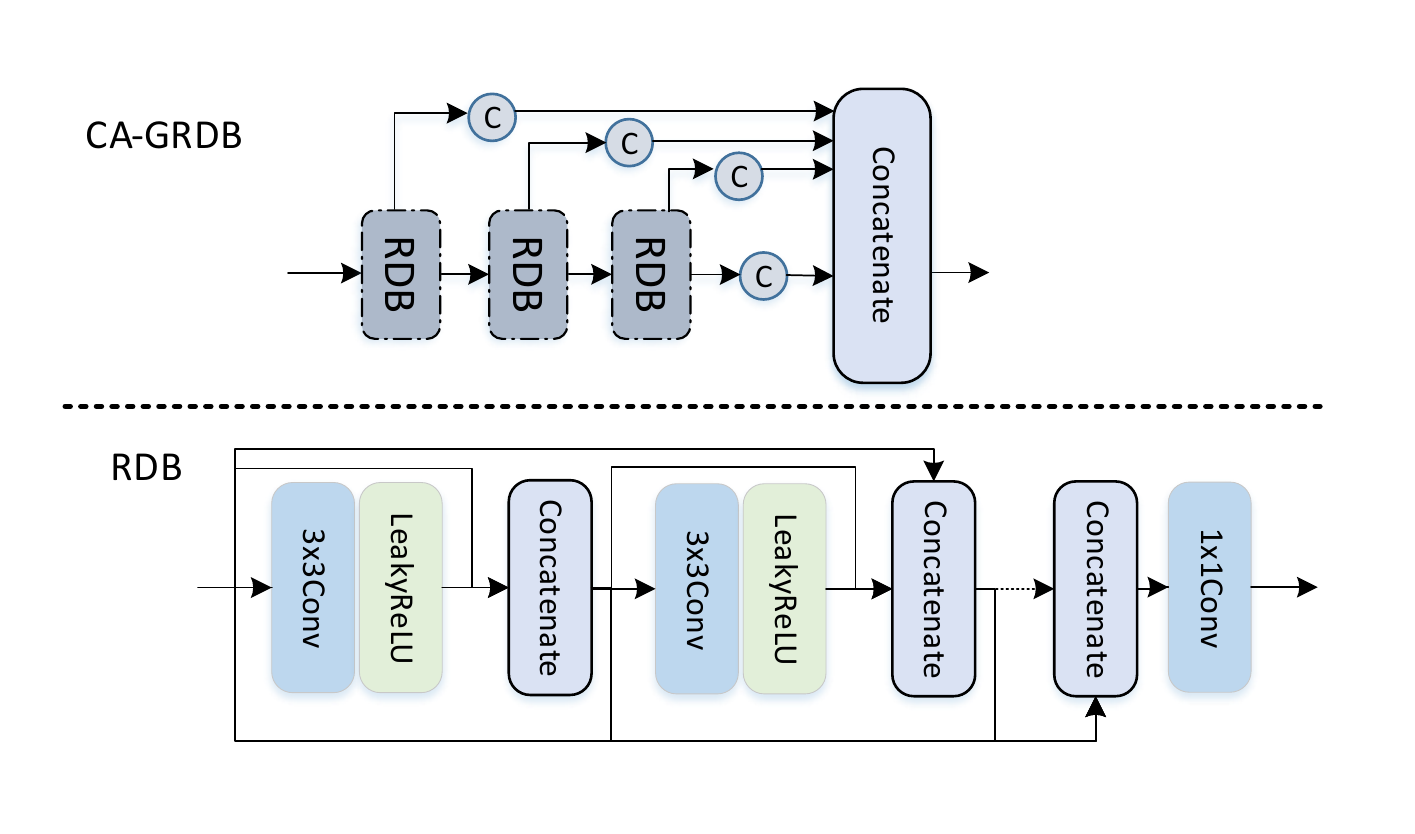}
	\caption{Components of CA-GRDB and RDB, where C denotes channel attention module from \cite{zhang2018image} and the convolutional layers in RDB are from \cite{kim2019grdn}.}
	\label{fig:CA-GRDB}
\end{figure}

\subsection{Main Body: Three Branches}
Since the degradation of DSLR camera is complex and contains multiple distortions such as blur and noise, we need to improve the network representation ability. After HFE, we hence propose the main body with three same branches to handle each frequency feature map separately. 
Inspired by \cite{zhang2018image} and \cite{kim2019grdn}, we introduce channel attention mechanism into the grouped residual dense block (GRDB) termed as CA-GRDB to adaptively aggregate those residual dense blocks (RDBs) \cite{zhang2018residual} with different attention weights. We cascade four CA-GRDBs in each branch where each CA-GRDB consists of three RDBs, resulting 12 RDBs of each branch shown in Fig. \ref{fig:framework}. 

Then the specific structure of CA-GRDB and RDB is shown in Fig. \ref{fig:CA-GRDB}. In CA-GRDB, we cascade three RDBs and apply channel attention after each RDB. Then the feature maps from channel attention are concatenated together as fusing operation. As for the RDB, we replace ReLU with LeakyReLU to avoid dying ReLU problem \cite{lu2019dying}. And each RDB is constructed by 8 layers with the form 3x3 Conv-LeakyReLU-Concat via dense connection and residual connection.



\begin{figure}[htp]
	\centering
	\includegraphics[width=\linewidth]{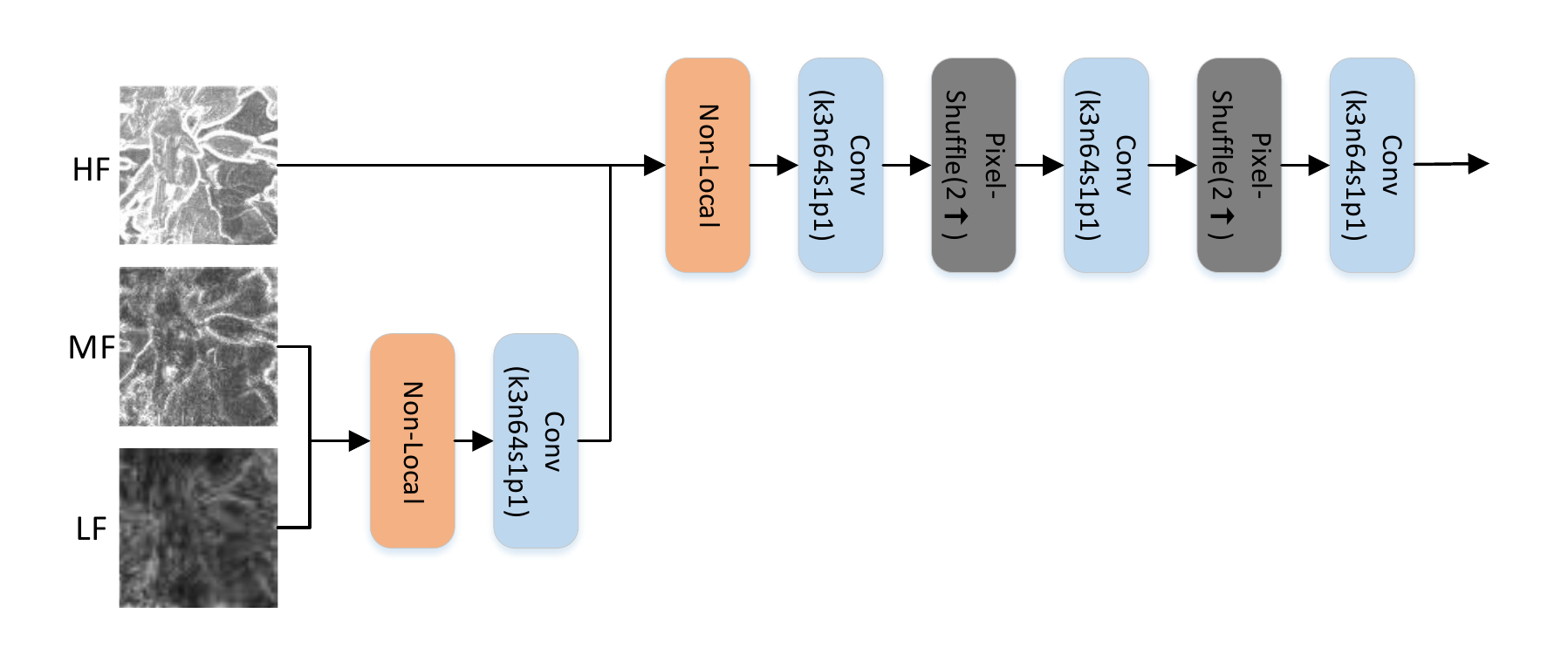}
	\caption{Fusion module, which integrate non-local attention \cite{Wang_2018_CVPR}, pixel-shuffle layer \cite{Lim2017Enhanced} and convolution layer to capture the correlation between different frequency features. Each convolutional layer is denoted with corresponding kernel size(k), number of feature maps(n), stride(s) and padding size (p). ($2\uparrow$) means 2x upsampling operation using pixel-shuffle layer.}
	\label{fig:fusion}
\end{figure}
\subsection{Fusion Module}
The fusion module is shown in Fig. \ref{fig:fusion}. Like \cite{Wang_2018_CVPR} claimed, Non-Local module (NL) tends to capture the long-term dependencies of features to compensate the lost information due to the local neighborhood filtering in convolutional layers.
To fuse the features of different branches, we utilize non-local attention and convolution layer to further capture the correlation between different frequency features. Then we exploit pixel-shuffle layer used in \cite{Lim2017Enhanced} and convolutional layer twice to upsample the frequency feature to the original image size, which is the generated HR image. Notice that we employ 2x upsampling operation using pixel-shuffle layer and we detail each convlutional layer with corresponding kernel size (k), number of feature maps (n), stride (s) and padding size (p) in Fig. \ref{fig:fusion}.

\subsection{Loss Functions}
We first employ pixel-wise $L1$ loss to measure the reconstruction error and then exploit $MSE$ loss to fine tune our model. And the overall loss function can be represented as follows,
\begin{equation}
    \mathcal{L_{\text{ALL}}}=\mathcal{L_\text{L1}}+\mathcal{L_{\text{MSE}}}
\end{equation}


\section{Experiments}
In this section, we will illustrate the RealSR dataset and evaluation metrics used in our experiments. And we will describe the implementation details of our training and testing. We compare our FAN with seven state-of-the-art methods qualitatively and quantitatively. To further verify the effectiveness of our each FAN module, We also conduct several ablation studies. 

\subsection{Dataset}
We train our FAN model based on 19,000 LR-HR pairs of training images provided by AIM 2020 Real Image Super-Resolution Challenge-Track3, where the paired LR-HR images are captured by various camera in practical scenarios.
In addition, the validation dataset contains 20 pairs of images. Although the test dataset provides 60 pairs of LR-HR images, the HR images are not available in the test stage. 
Therefore, we test our FAN model on validation dataset and compare with other state-of-the-art methods in this work. According to the released final results, we achieved the fourth place on the X4 track with PSNR of 31.1735 and SSIM of 0.8728 on test dataset.

\subsection{Evaluation Metrics}
\noindent(1) \textbf{Peak Signal-to-Noise Ratio (PSNR)}: PSNR is one of the most widely used full-reference quality metrics. It measures the ratio between the maximum possible value (power) of a signal and the power of distorting noise that affects the quality of its representation. In other words, PSNR metric reflects the intensity differences between the real HR image and the recovered HR image. A higher PSNR score means that the intensity of two images is more close.
Mathematically, it is high related to MSE loss mentioned above:
\begin{equation}
    PSNR = 10log_{10}(\frac{R^{2}}{MSE})\quad
\end{equation}
where $R^{2}$ denotes the maximum fluctuation in the input image data type.

\noindent(2) \textbf{The Structural SIMilarity Index (SSIM) \cite{wang2004image}}: SSIM evaluate the image quality based on luminance, contrast and structure from the perspective of image formation. We apply it to compute the perceptual distance between the real HR image and the recovered image. A higher SSIM score means that the luminance, contrast and structure of two images are more similar.
Mathematically,
\begin{equation}
\begin{aligned}
&   \mathcal{L_{\text{SSIM}}}=[l(\hat{I},I)^{\alpha}\cdot{[c(\hat{I},I)]^{\beta}}\cdot{[s(\hat{I},I)]^{\gamma}}\\
& l(\hat{I},I)^{\alpha}=\frac{2\mu_{\hat{I}}\mu_{I}+C_{1}}{{\mu_{\hat{I}}^{2}+{\mu_{I}}^{2}+C_{1}}}\quad\\
& c(\hat{I},I)]^{\beta}=\frac{2\sigma_{\hat{I}}\sigma_{I}+C_{2}}{\sigma_{2}^{2}+\sigma_{I}^{2}+C_{2}}\quad\\
& s(\hat{I},I)]^{\gamma}=\frac{\sigma_{\hat{I}I}+C_{3}}{\sigma_{\hat{I}}\sigma_{I}+C_{3}}\quad\\
\end{aligned}
\end{equation}
where $\mu_{\hat{I}}$, $\mu_{I}$,$\sigma_{\hat{I}}$,$\sigma_{I}$ denote mean and standard deviation of the image $\hat{I}_{HR}$, $I_{HR}$ respectively. $C_1$, $C_2$, $C_3$ are three constants to avoid instability. $\alpha$, $\beta$, $\gamma$ are hyper-parameters to control the relative importance.

\subsection{Implementation Details}
Our FAN is implemented based on PyTorch framework with four NVIDIA 1080Ti GPUs. To avoid undesirable over-fitting behaviors due to the limited data, we use data augmentation in the process of training such as randomly cropping, flipping and rotation. The number of mini-batches was set as 8. We use Adam optimizer with a initial learning rate of 0.0001 which will decay by a factor 0.5 every 30 epochs. We spent almost 48 hours to train our FAN model. 

For testing, we first crop the LR image into several $196\times196$ small patches and then fed them into our FAN model to obtain the HR patches. Finally we convert those patches into a complete HR image. Our model requires 0.19s on a single NVIDIA 1080-Ti GPU with an LR image of $196\times196$ for testing. And we all use the 130 epochs.


\subsection{Comparison with the state-of-the-art methods}
We compare our FAN model with seven state-of-the-art methods in same test settings. The first one is a traditional method, i.e, bicubic upsampling. And the others are CNN-based methods which are retrained on the RealSR x4 dataset for fairness. DBPN \cite{haris2018deep} exploits iterative up and downsampling 
layers to self-correct features at each stage by error feedback mechanism. EDSR \cite{Lim2017Enhanced} modifies the conventional residual networks by removing some unnecessary modules and get significant performance improvement. RCAN \cite{zhang2018image} proposes the channel attention to capture the dependencies among deep residual feature maps. RDN \cite{zhang2018residual} conducts a novel residual dense network to extract feature maps. SAN \cite{dai2019second} incorporates the non-local attention mechanism to capture long-range spatial contextual information. VDSR \cite{Kim_2016_CVPR} extends the depth of neural network which highly improves the accuracy.
\begin{table}[htp]
\caption{Quantitative results on RealSR x4 dataset. We compare our FAN with the state-of-the-art methods in terms of PSNR and SSIM.}
\centering
\begin{tabular}{c|c|c|c|c|c|c|c|c}
\hline
Method & VDSR & EDSR & RDN & DBPN & RCAN & SAN & FAN(ours) & FAN+  \\ \hline
PSNR & 28.966 & 29.522 & 29.790 & 29.305 & 30.002 & 30.112 & \textbf{30.598} & \textbf{30.719}\\ \hline
SSIM & 0.8322 & 0.8444 & 0.8486 & 0.8388 & 0.8529 & 0.8557 & \textbf{0.8603} & \textbf{0.8621}\\ \hline
\end{tabular}
\label{compare}
\end{table}

Table \ref{compare} shows the comparisons of our FAN with the previous methods on RealSR x4 dataset. Compared to other methods, our FAN with self-ensemble ($\times8$) termed as \textbf{FAN+} containing flipping ($\times4$) and rotation ($\times4$). FAN+ both performs the best scores in PSNR and SSIM, leading to a $0.607$ dB and a $0.0064$ dB increase respectively compared with SAN \cite{dai2019second}. Even without self-ensemble, termed as \textbf{FAN(ours)}, can also achieve $0.486$ dB and $0.0046$ dB increase.

\begin{figure}[htp]
	\centering
	\includegraphics[width=\linewidth]{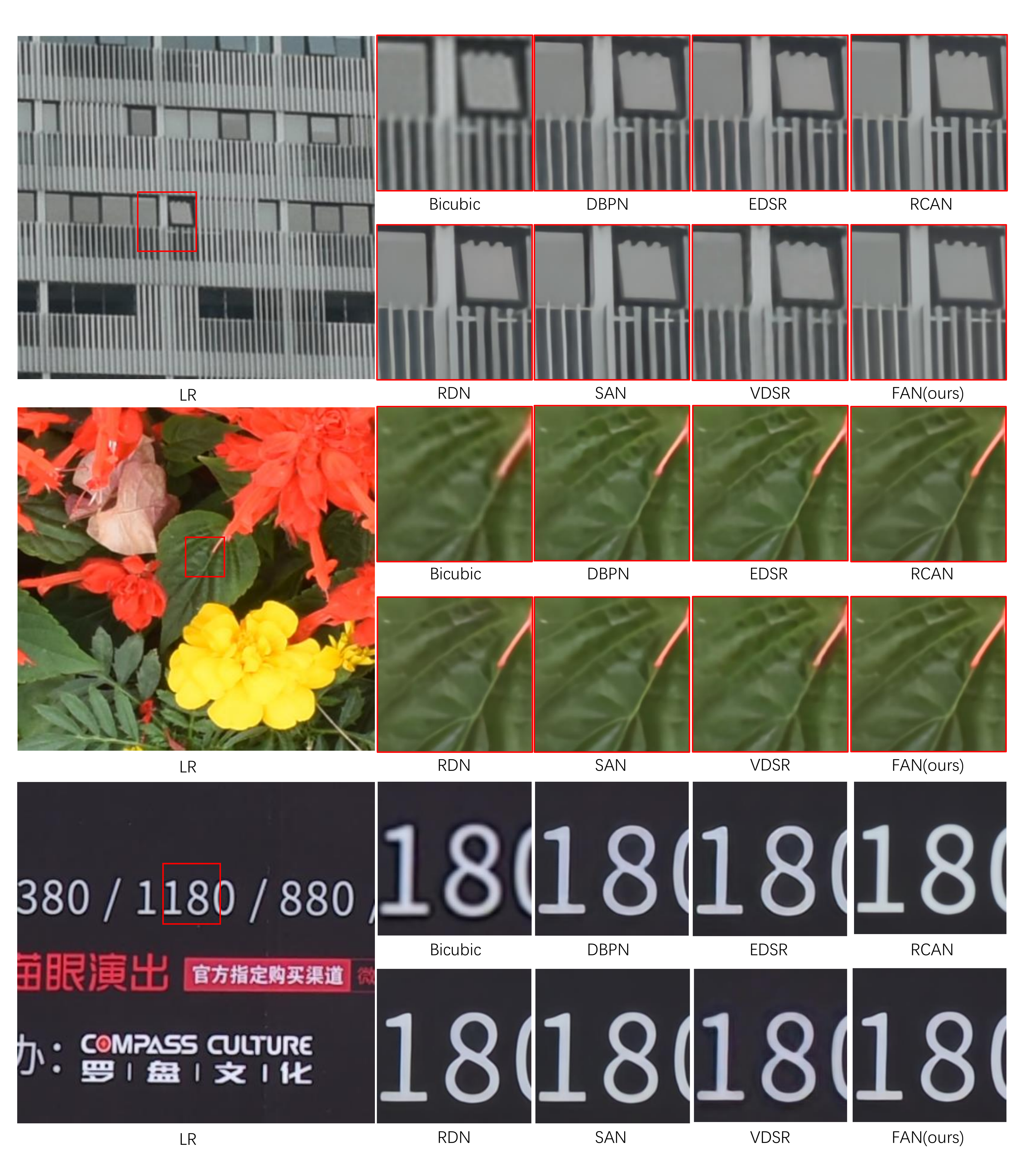}
	\caption{The performance comparison of our FAN with the state-of-the-art methods performed on RealSR x4 dataset. The state-of-the-art methods are bicubic upsampling, DBPN \cite{haris2018deep}, EDSR \cite{Lim2017Enhanced}, RCAN \cite{zhang2018image}, RDN \cite{zhang2018residual}, SAN \cite{dai2019second} and VDSR \cite{Kim_2016_CVPR} respectively.}
	\label{fig:results}
\end{figure}

The qualitatively comparison is shown in Fig. \ref{fig:results}. The traditional bicubic upsamping produces the most blurry results. 
And compared with the other cnn-based methods, our FAN model can generate sharper and clearer HR images without obvious artifacts or distortions which are caused by the enlarged DSLR camera lens in the RealSR dataset. For example, the building fences and leaf veins are more clearer and not anamorphic. All in all, our FAN model can solve the RealSR problem well while the previous SISR methods can not generalize to this problem.


\begin{figure}[htp]
	\centering
	\includegraphics[width=\linewidth]{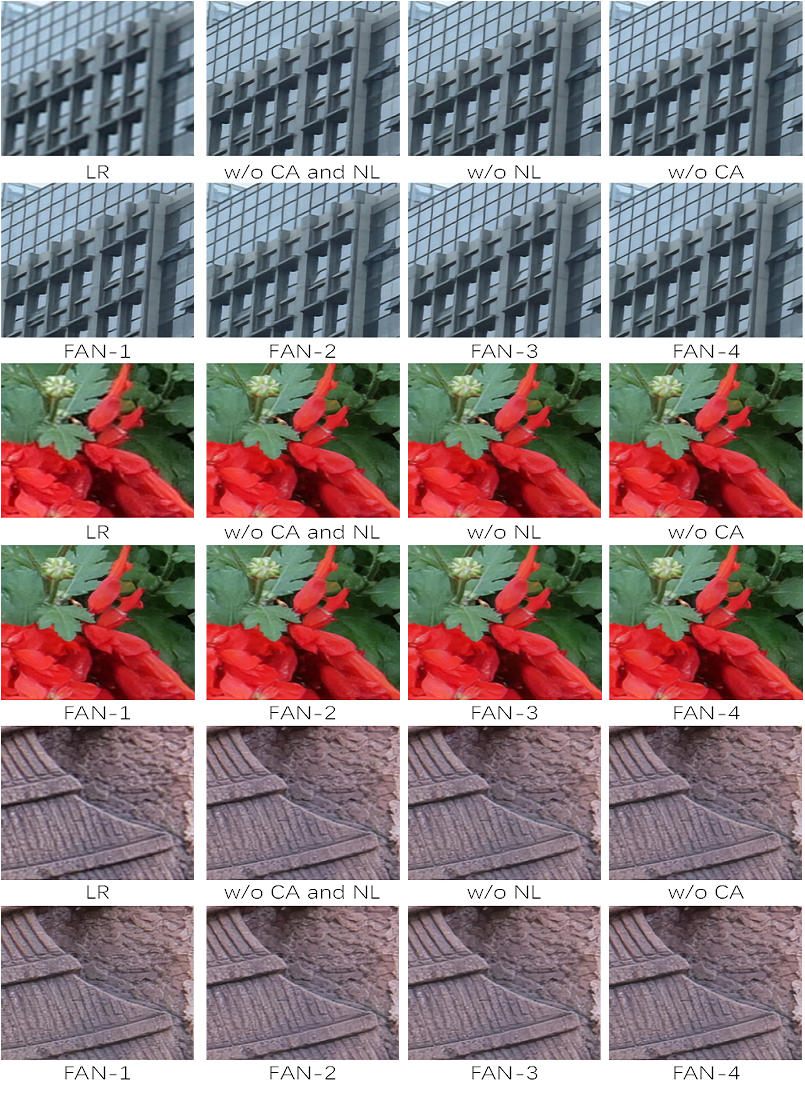}
	\caption{The subjective performance comparisons in ablation studies of FAN.}
	\label{fig:ablation}
\end{figure}

\subsection{Ablation study}
To further demonstrate the effectiveness of our each module, we conduct several ablation studies to investigate the influence of the multi-frequency branches, channel attention for CA-GRDB, non-local module, channel attention mechanism and the number of RDBs in our FAN.The subjective performance comparisons in ablation study are shown in Fig. \ref{fig:ablation}.

\noindent \textbf{Dataset}
Since the ground truth HR images of validation dataset are not released, we randomly split the training dataset into two parts by the ratio 7:3 for training and testing respectively. Notice that this dataset setting is only for ablation study. 

\noindent \textbf{Multi-frequency branches}
We set the number of branches from 1 to 4 to figure out the effects of different branches, denoted as FAN-1, FAN-2, FAN-3, FAN-4 shown in Table \ref{tab:branches}. with the number of branches increasing, we can observe the improvement of performance in general. However, the gain gets smaller and the number of parameters increase significantly when the number of branches is more than 3. Therefore, we select 3 as a trade-off to balance the complexity of network and performance.

\noindent \textbf{Channel attention(CA) and Non-local module(NL)}
We exploit CA to model the feature inter-dependencies, which enforces our model to concentrate on more informative features. Like \cite{Wang_2018_CVPR} claimed, NL tends to capture the long-term dependencies of features to compensate the lost information due to the local neighborhood filtering in convolutional layers.
From Table \ref{tab:ablation}, we can see that whether removing CA (w/o CA) or removing NL (w/o NL) results in the drop of performance. At the same time, adding CA and NL both would achieve best performance while the number of parameters only has slightly increasing.


\begin{table}[htp]
	\centering
	\caption{Comparisons between different number of branches in FAN. Tested on partitioned dataset, which is different from validation dataset.}
	\setlength{\tabcolsep}{0.25mm}{
		\begin{tabular}{c|c|c|c|c}
			\hline
			Branches & FAN-1 & FAN-2 & FAN-3 & FAN-4 \\ \hline
			Parameters (MB) & 5.81      & 27.95    & 50.30    & 72.74\\ \hline
			PSNR (dB)  & 28.5210   & 28.8029  & 28.9464   & 28.9598        \\ \hline
			SSIM   & 0.8153   & 0.8194  & 0.8231   & 0.8233             \\ \hline
		\end{tabular}
	}
	\label{tab:branches}
\end{table}
\begin{table}[!htp]
	\centering
	\caption{Ablation study on different combinations of channel attention and non-local attention. Tested on partitioned dataset.}
	\setlength{\tabcolsep}{4mm}{
		\begin{tabular}{cc|ccc}
			\hline
			CA. & NL. & Parameters(MB) & PSNR(dB) & SSIM \\ \hline
			$\checkmark$& $\checkmark$   & 50.30 &28.9464 & 0.8231 \\
			$\times$& $\checkmark$  &50.28 &28.8073 & 0.8195\\
			$\checkmark$&$\times$   &50.12 &28.7931 & 0.8194 \\
			$\times$&$\times$   &50.10 &28.7571 & 0.8190 \\
			 \hline
		\end{tabular}
	}
	\label{tab:ablation}
\end{table}

\section{Conclusions}
In this paper, we proposed FAN, a frequency aggregation network, to solve the real-world image super-resolution problem which suffers from more complicated degradation process rather than simple bicubic downsampling. Our model is designed in an end-to-end manner, in which HFE extract three feature maps contained different frequency information firstly. And then pass them into CA-GRDB module independently to output corresponding residual dense feature map in each frequency. Finally, we apply fusion module to integrate them to recover the HR image. We validate the effectiveness of our FAN in RealSR Dataset with extensive experiments. In addition, our model performs well on the AIM 2020 challenge which came the fourth place on the X4 track with PSNR of 31.1735 and SSIM of 0.8728.

\section*{Acknowledgement} 
This work was supported in part by NSFC under Grant U1908209, 61632001 and the National Key Research and Development Program of China 2018AAA0101400.



\clearpage
%
%
\bibliographystyle{splncs04}

\bibliography{egbib}
\end{document}